\documentclass[runningheads]{llncs}
\usepackage{graphicx}
\usepackage{caption}
\begin{document}
\title{Quality Management of Machine Learning Systems}

\author{P. Santhanam}
\institute{IBM Research AI, T.J.Watson Research Center \\ 
Yorktown Heights, New York \\
\email{pasanth@us.ibm.com}}
\maketitle

\begin{abstract}
In the past decade, Artificial Intelligence (AI) has become a part of our daily lives due to major advances in Machine Learning (ML) techniques.  In spite of an explosive growth in the raw AI technology and in consumer facing applications on the internet, its adoption in business applications has conspicuously lagged behind. For business/mission-critical systems, serious concerns about reliability and maintainability of AI applications remain.  Due to the statistical nature of the output, software `defects' are not well defined.  Consequently, many traditional quality management techniques such as program debugging, static code analysis, functional testing, etc. have to be reevaluated. Beyond the correctness of an AI model, many other new quality attributes, such as fairness, robustness, explainability, transparency, etc. become important in delivering an AI system.  The purpose of this paper is to present a view of a holistic quality management framework for ML applications based on the current advances and identify new areas of software engineering research to achieve a more trustworthy AI.

\end{abstract}

\section{Introduction}

According to the 2019 AI Index report [1], hundreds of papers are published on AI technology every day! In 18 months, the time required to train a large image classification system on the cloud infrastructure has fallen from about 3 hours to about 88 seconds! In 2019, global private AI investment was over \$70B. In 2018, the State of California licensed testing of more than 500 autonomous vehicles, which drove over 2 million miles. When it comes to real AI applications, many companies on the internet use the latest machine learning techniques to perform various consumer facing tasks, such as, answer questions, recognize images, recommend products, translate content, etc.  Not a day goes by when there is not a news report of a new application of machine learning to some new domain. All these trends point to an explosion of AI related technology all around.    

Interestingly, the adoption of AI in the enterprise for business critical applications has lagged considerably behind. Recent analysts reports from various sources [2] indicate at most 20-40\% success rate for the adoption of AI to create business value. This supports the assertion that moving AI from a proof-of-concept to real business solution is not a trivial exercise. Some common reasons cited for this result are:

\begin{itemize}
\item Insufficient alignment of business goals and processes to the AI technology (akin to the challenges of introducing information technology in the 1990's).
\item Lack of data strategy (i.e. ``There is no AI without IA (Information Architecture)'') 
\item Shortage of skilled people who can combine domain knowledge and the relevant AI technology.
\item Unique concerns about AI (e.g. model transparency, explainability, fairness/bias, reliability, safety, maintenance, etc.)
\item Need for better engineering infrastructure for data and model provenance.
\end{itemize}

As the application of AI moves to business/mission critical tasks with more severe consequences, the need for a rigorous quality management framework becomes critical.  It is bound to be very different from the practices and processes that have been in place for IT projects over many decades. The goal of this paper is to provide an overview of such a framework built upon tools and methodology available today and identify gaps for new software engineering research. The focus of this paper is on AI systems implemented using machine learning.  A popular ML technique is the use of Deep Neural Networks (DNN). This paper uses AI and ML interchangeably.

\section{AI is Software}  

In general, the use of an AI component has one of three goals. (i) automate an existing task performed by a human e.g. Support Bots (ii) improve the efficiency of an existing task e.g. language translation (iii) perform a new task e.g. a recommender system. The invocation of the AI is through traditional interfaces (e.g. REST based microservices). In this respect, it is just another software component, albeit with some special attributes. Thus, from the system or software management point of view, it has all the same expectations as any other software component.   Figure 1 shows the recommended system and software quality models and attributes from the ISO/IEC 25010 process standard [3]. Reference [4] gives an accessible overview of the relevant attributes. Even though the specific interpretation may have to be refined for the purpose of AI components, the utility of the basic structure is immediately evident. The quality attributes in use (in the left column) i.e. effectiveness, efficiency, satisfaction, risk and context coverage do represent the relevant dimensions for consideration.  Inclusion of `Trust' under the `Satisfaction' category is fortuitous in hindsight, since it has taken a more profound meaning for AI components.  The product quality attributes on the right are essential for product owners.  Notably, the common metric used by AI algorithm owners is accuracy which relates to `Functional Correctness' in Figure 1 and it is only one of the more than two dozen attributes in the ISO standard. \emph{It is important to  evaluate an AI component against these attributes to understand the requirements they place on the use of an AI component in a software system.}

%--------------------------------------
\begin{figure*}[h!]
\centering
\includegraphics[width=\textwidth,scale=0.5]{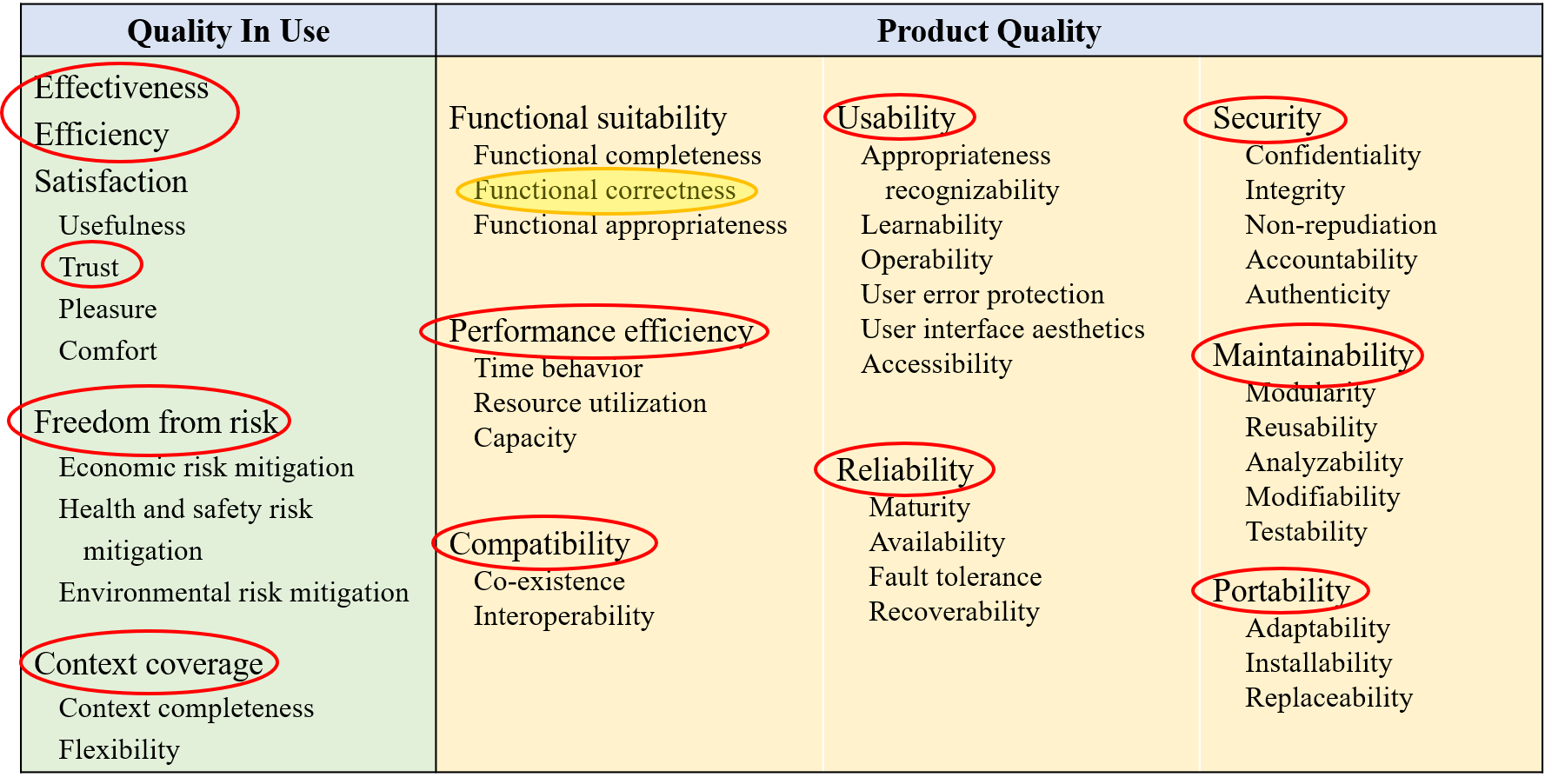}
\caption{ISO/IEC 25010- System and Software Quality Models [3,4]}
\label{fig:ISO}
\end{figure*}
%--------------------------------------

\subsection{Traditional Software Quality Management}

The engineering practices for managing software quality go back many decades [5], and McConnell [6] gives a more recent view of practices and tools. A key assumption is that \textbf{expected functional behavior (i.e. linking inputs and outputs) of components is documented (or understood) at design time}, even if design evolves during a project. Central to quality management is the definition of a \textbf{defect} (aka bug) as the software behavior not meeting the expectation. Defects can be opened during any of the development activities [7], namely, Design or Code Review, Unit Test (white box), Function Test (black box), System Test (integration), and during DevOps or Operations. 

There are seven aspects to managing quality processes in a software project. (i) \textbf{Requirements management} that evaluates the incoming (and often changing) requirements and selects the requirements for inclusion in the upcoming software release(s).  (ii) \textbf{Defect management} which includes the process of opening, debugging, fixing, closing and counting defects across the life cycle.  (iii) \textbf{Change management}, which relates to versioning and tracking of changes to code and documentation during the life cycle. (iv) \textbf{Test Management}, consisting of Test design, Test creation, Test execution and Evaluation metrics (e.g. test effectiveness, defect density, code or functional coverage, etc.) This applies across all levels of testing i.e. unit testing (white box), function testing (black box), etc. (v) \textbf{Dev/Op processes} that manage the promotion of code from development to operations with the necessary processes and automation (e.g. regression tests) to validate deployment quality and support the run-time environment. (vi) \textbf{Operations management} that collects incident reports during operations and provides a mechanism for support teams to diagnose and resolve them expediently, which may involve the original engineers who created the relevant code.  
(vii) \textbf{Project management} that brings these six different aspects into a cohesive decision support system for risk management via dashboards.

\subsection{Machine Learning Systems}

A recent paper [8] discusses the engineering challenges in building reliable ML systems. At a high level, any AI solution based on machine learning technology is an interplay between three factors i.e. Data, Domain context and the AI Algorithms (Figure 2). The solution quality is determined by the algorithms that learn from the training data to create outputs that make sense to the humans in the specific domain context/application area. These three factors together define the necessary conditions. If any one is missing, the resulting ML system is likely to fail. Understanding this is critical to assess the expected business value of an AI system.
%--------------------------------------
\begin{figure}[h]
\centering
\includegraphics[scale=0.5]{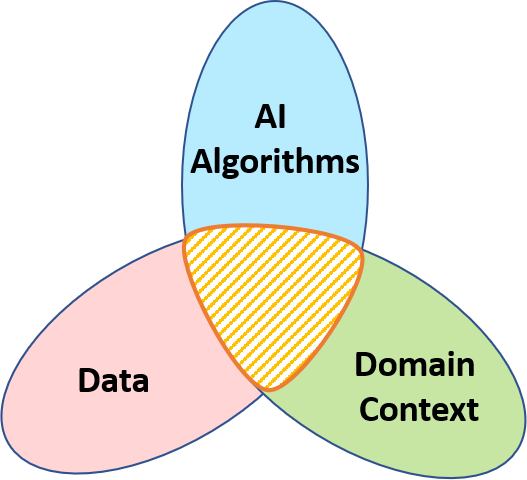}
\caption{Critical Success Factors for AI applications}
\label{fig:triad}
\end{figure}
%--------------------------------------

%--------------------------------------
\begin{table*}[h!]
\centering
\includegraphics[width=\textwidth,scale=0.5]{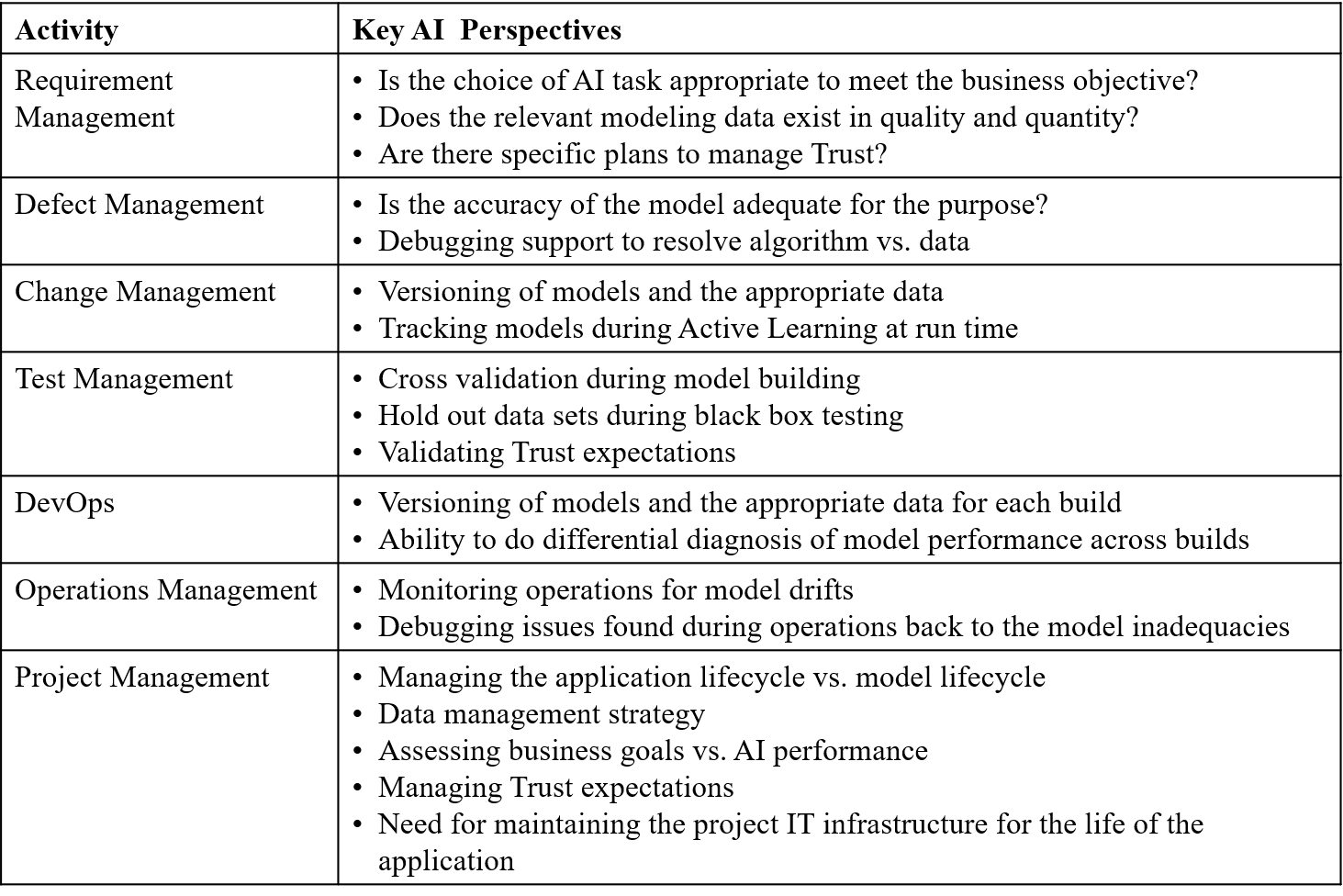}
\caption{Key perspectives in quality management of AI projects}
\label{tab:QM4AI}
\end{table*}
%--------------------------------------

In supervised machine learning, the modeling data consists of large number of inputs and the corresponding outputs (called labels). The main goal of the  ML algorithm is to find the best approximation to the function that maps the inputs to the output(s). During inference, the output is the prediction that can be a continuous variable (e.g. price of a stock or tomorrow's temperature) or a label/class (e.g. predicting the next word in a search engine, identifying an image as a cat, etc.) These ML functions do not have a requirement/specification document at the design time; they are just derived from the modeling data.  Model outputs are statistical and not deterministic.  Consequently, there is no simple way to define a defect! As an example, an image recognition algorithm may identify a `cat' as a `dog' for some of the instances and this is allowed by the statistical uncertainties in the algorithm.  Debugging is complicated since the problem can be in the model and/or the data.  No guarantees or explanations are provided on the exact functional operations of the model.  Traditional testing will not work, since there is no description of the expected behavior at design time.  In addition, ML Model behavior can drift over time during deployment due to previously unseen data. If the model is learning continuously during deployment, new patterns of relationships in data can emerge, unknown to the model owners. 

The potential breadth of AI applications invokes serious social concerns as evidenced by various government initiatives such as the European Commission Ethics Guidelines for Trustworthy AI [9] and AI Ethics Principles for the US Department of Defense [10].  Any AI quality management framework has to address these concerns.  Table 1 describes the changes to the quality processes discussed in section 2.1 due to the inclusion of an AI component. It is clear that \textbf{every one of the traditional quality management activities is affected.}  Due to the nature of the AI applications, the quality management is a required and never ending activity, as long as the application is in use.   

\section{A Quality Management Framework for ML systems}

The purpose of this section is to identify the key components that are needed to have an adequate quality management framework for successful delivery of reliable ML systems. Some of the components have prototype technology already available for use, but while others need more experimentation and new research. This discussion leverages some of the concepts from references [11-14].
%--------------------------------------
\begin{figure*}[h!]
\centering
\includegraphics[width=\textwidth,scale=0.5]{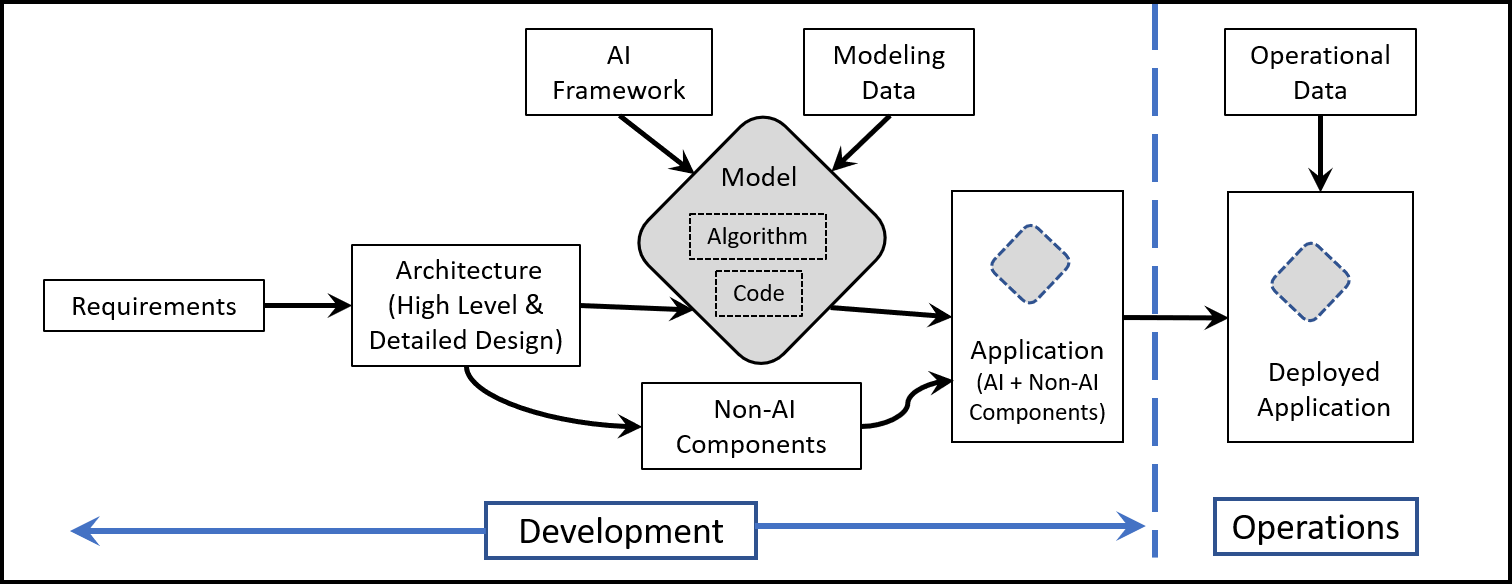}
\caption{Key artifacts in the creation of an application with one ML component implemented in the model (gray rounded square). For simplicity, iterative process steps are not shown and only the AI component properties are emphasized.}
\label{fig:AI-Artifacts}
\end{figure*}
%--------------------------------------

Figure 3 shows the key artifacts in the development of a ML application.  For simplicity, we emphasize AI specific artifacts.  All software projects start with a set of functional and non-functional requirements, mostly in natural language documents.  Software architecture consists of high level and detailed designs typically captured in various diagrams [6].  Detailed designs identify the specific functions needed for the implementation. One or more of the functions can be implemented by data scientists using ML models.  Modeling data refers to the data available for training and validation.  The data scientists either reuse algorithms and other components from popular AI frameworks (e.g. TensorFlow, PyTorch, etc.) or write the modeling code for the first time. Model gets integrated with other non-AI components (e.g. user interface, traditional analytics, reporting, etc.) and IT components (e.g. access control) to create the business application which is subsequently deployed to operations.

\subsection{Where are the bugs?}

The defects in application requirements and design are nothing new in software development, but ML introduces some twists in the assessment of the right task for the automation with AI, based on business expectations on quality (discussed above) and operational performance. Incorrect choice of algorithm for the chosen task is a common source for quality concerns. The programming errors in the implementation of the algorithm is also not a new problem. Examples are incorrect API calls, syntax errors, incorrect model parameter, etc. These can be addressed with additional support from the frameworks, much like Eclipse for Java.     

As in all projects that involve statistics, the quality and quantity of the available data are major concerns.  An example of raw data quality is ''expectation mismatch''(i.e.incorrect data type for feature values, Non-boolean value for boolean feature type, etc.)  
The more subtle data problems relate to noisy labels in the modeling data and issues with data distributions (e.g. the fraction of examples containing a feature is too small, data distribution is different for training and validation data sets, etc.) Data problems also resurface in production when the operational data profile does not match the data used during model development or due to unexpected emergent behavior.  

Due to the extensive use of open source machine learning frameworks (i.e. TensorFlow, CNTK, Keras, Theano, PyTorch, etc.) and associated libraries, they become additional sources of bugs [15,16]. Testing a ML library for a specific algorithm (e.g. convolutional neural network, recurrent neural network, etc.) will require implementation of the same algorithm in different ML libraries and luckily this is not a problem with common algorithms and popular libraries. Examples of bugs in frameworks are: not backward compatible API versions, unaligned tensors, pooling scheme inconsistency, etc.

Then, there are the bugs in the model itself, as evidenced by an unexpected output for the given input (e.g. a cat's image identified as a dog) , through various modeling errors such as overfitting, unoptimized hyper parameters, wrong neural net architecture, etc.
Once the model is integrated into the business application, the defects in the ML components get mixed up with the traditional software defects in the other components. The overall quality of the application is obviously dependent on all the contributing software components, such as user interface, back end management, etc.    

\subsection{Quality Improvement Tasks for ML systems}

This section describes the suggested tasks to find defects in the artifacts described in Section 3.1 and resolve them. These are traditional activities modified to reflect the inclusion of the ML component in the application.  \emph{Due to space limitations, reference to any specific technique or tool is meant to provide an example, rather than an exhaustive list}. Quality improvement tasks that address the unique aspects of assessing `Trust' in ML systems are described in Section 3.3. 

\subsubsection{Manual Inspection}
With the support of tools [17] manual inspection is still an effective way to find defects in requirements, architecture, algorithms and code. Techniques, such as pair programming [6] have proven very useful in practice. 

\subsubsection{Static Analysis}
Application of static analysis to find defects in software programs is a very mature field, dating back to many decades [18]. There have been recent examples of applying this technique to machine learning code [19].  Many development environments support basic syntax checking, when the code is being written.

\subsubsection{White Box Testing}
Traditional white box testing [20] leverages the knowledge of the program structure to execute the program in ways to achieve the desired coverage e.g. branch coverage, statement coverage, etc. to locate defects.  Similarly, a data scientist can use the detailed knowledge of a neural network behavior in the model building process to apply various coverage criteria to the network to find the defects in the model. This has led to concepts such as neuron coverage [21], novel test criteria that are tailored to structural features of DNNs and their semantics [22], the condition-decision relationships between adjacent layers and the combinations of values of neurons in the same layer [23], mutation techniques on source code, data and models [24] and combinatorial test design consisting of neuron pairs in the layers and the neuron-activation configurations[25].  These techniques demonstrate various ways to expose incorrect behavior of the network while being mindful of the computational cost of test generation itself.    

\subsubsection{Black Box Testing}
Traditional black box testing (or functional testing) [20] focuses on detecting defects in the expected external behavior of the software component by carefully manipulating the input space.  For ML models, it is important that test data represents the business requirements in terms of data values and distributions and was not used during the model creation process.  Key goal of black box testing is to evaluate if the model generalizes adequately for previously unseen data or suggest a model rework, if not suitable. These considerations also apply for system integration tests.

\subsubsection{Data Assessment \& Testing}
There are several techniques and tools to check the quality of the modeling data during development.  Breck et al.[26] present a highly scalable data validation system, designed to detect data anomalies (e.g. unexpected patterns, schema-free data, etc.) in the machine learning pipelines. Barash et al.[27] use combinatorial design methodology to define the space of business requirements and map it to the ML solution data, and use the notion of data slices to identify uncovered requirements, under-performing slices, or suggest the need for additional training data.  This is also an example of using data slicing for black box testing.

\subsubsection{Application Monitoring}
Application monitoring during operations is a critical activity in ML applications since the model performance can change over time due to previously unseen pattern in the operational data or emergent behavior not expected in the model building process.  Breck et al [26] also describe techniques to detect feature skew by doing a key-join between corresponding batches of training and operational data followed by a feature wise comparison. Distribution skew between training data and serving data is detected by distance measures.  Raz et al.[28] discuss a novel approach, solely based on a classifier suggested labels and its confidence in them, for alerting on data distribution or feature space changes that are likely to cause data drift. This has two distinct benefits viz. no model input data is required and does not require labeling of data in production. In addition to the detecting any degradation of model performance, there need to be processes in place to correct the behavior as and when it occurs. There are examples of commercial offerings to perform this task [29]. 

\subsubsection{Debugging}
Debugging is often the most under-appreciated activity in software development that takes considerable skill and effort in reality. As noted in [7], debugging typically happens during three different stages in software life cycle, and the level of granularity of the analysis required for locating the defect differs in these three. First stage is during the model building process by the data scientist who has access to the details of the model. Here, there are two classes of errors that need debugging. (a) raw errors resulting in the execution of the model code in the development environment during the process of model creation. The development frameworks can provide support for debugging this class of problems. (b) Model executes successfully, but the overall performance of the model output is not adequate for the chosen task or if the model output does not meet the expectation for specific input instances, it is necessary to find the reason for these behaviors.  There could be many causes, including bad choice of the algorithm, inadequate tuning of the parameters, quality and quantity of the modeling data, etc. [30, 31]. Some care is also  needed in providing model debugging information to the user to avoid exposure of system details, susceptible for adversarial attacks.  

The second stage for debugging is during the later black box testing activities in development when an unexpected behavior is encountered. A certain amount of debugging of the test execution is necessary to conclude that the AI model is the cause of the unexpected behavior. Once that is confirmed, debugging of the model follows the same process as described above. Third stage is during operations, when the application is being put to real use. Any unexpected behavior here can be the result of changes in the computing environment relative to development or due to new patterns in the operational data not previously seen the modeling data. Techniques discussed in [26, 28,29] can be used to address the model drift problems. 

\subsection{AI Trust Assessment}
Due to the black box nature of the ML models and their behavior being decided by modeling data, trust in model outputs has become an important consideration in business applications. This section deals with four specific aspects trust. 

\subsubsection{Explainability}
In many business critical applications, the outputs of the black box ML models also require explanations to meet the business objectives. There are many motivations for explanations [32] and it is important to know the need so that the appropriate approach can be used. There are examples of open source packages for implementing explainability [33] in business applications.   

\subsubsection{Bias/Fairness}
Due to the potential sensitivity of the outputs of the ML models to biases inherent in the modeling data, there is a critical question of the fairness of the algorithms [34] in extracting the model from the data. There are examples of open source packages for understanding and mitigating biases [35] in business applications. 

\subsubsection{Robustness}
The owner of a ML application needs a strategy for defending against adversarial attacks. Xu et al.[36] provide a comprehensive summary of the adversarial attacks against ML models built using images, graphs and text and the countermeasures available.  Reference [37] describes an open-source software library, designed to help researchers and developers in creating novel defense techniques, and in deploying practical defenses of real-world AI systems. 

\subsubsection{Transparency}
Given the abundance of AI components (i.e. algorithms, services, libraries, frameworks) available from open source and commercial offerings, it makes sense for a company to reuse the available software component in its application. However, due to the concerns about the trust in the available component, the consumer of the component needs some detailed information about the component to manage the risk.  This need for transparency requires additional assessment. Key pieces of such information are captured in a FactSheet [38], which provides the technical and process background of the AI asset to the consumer.  

\subsection{Quality Metrics}
Due to the unique attributes of AI based systems (discussed in Section 2), there is a critical need to reevaluate the metrics for their quality management. This section discusses three aspects that highlight the need.  

\subsubsection{Defect management}
The lack of a clear definition of software defect in ML applications discussed in section 2.2 is a major problem in quality management. While the defects in the other artifacts can be captured unambiguously, the perceived errors in the model outputs are subject to the statistical uncertainties.  As a result, until a detailed debugging is performed to diagnose the reason for the unexpected behavior one cannot be certain that this is a bug. Hence the defect management tools have to allow this possibility with potentially more time assigned for the necessary investigation that may point to an inadequate training data set. 

\subsubsection{Model Evaluation}
Model evaluation is an important part of the ML application development. There are many metrics [39] that can be used and depending on the specific application domain. They need to be chosen and used carefully. In addition to these usual machine learning metrics, additional metrics specific to the trust topics discussed in section 3.3 (explainability, bias, robustness and transparency) are also necessary to support the business objectives and manage technical \& business risk. There is also recent work [40] to measure application-level key performance indicators to provide feedback to the AI model life cycle.

\subsubsection{Model Uncertainty}
In addition to the usual ML metrics [39], typically at the end of a DNN pipeline is a softmax activation (usually a sigmoid function) that estimates a confidence level for each output, expressed as a probability measure between 0 and 1. In reality, a high confidence level does not necessarily mean low uncertainty [41] and hence it is not reliable for decision support.  This is because DNN models do not have a way of calculating uncertainty by themselves. Gal and Ghahramani [41] have proposed a new method to estimate uncertainty in DNN outputs that approximates Bayesian models, while not requiring a high computing cost. Lakshminarayanan et al.[42] have demonstrated an alternate approach that is scalable and can be easily implemented.  Any mission critical ML system has to include such uncertainty measures.  

\section{Conclusions}
The purpose of this paper is to discuss a framework for a quality management system needed to support the inclusion of AI components in the building of business/mission critical applications.  It should be clear from the description above, that ML applications need a different mindset from the start of a project.  Since AI is really a software component, we need to apply the relevant software quality attributes to it. AI comes with some special quality attributes (fairness, explainability, etc.) which have to be integrated into the quality management methodology.   Various processes and tools to support the quality management of the application are in their infancy, mostly as prototypes from research projects.  In addition to the raw tooling needed for each task described in Sec.3, the integration of them across the life cycle to provide a holistic system is critical for wide scale use. Furthermore, ethical guidelines from governments [9, 10] require an engineering implementation to demonstrate adherence. This does not exist today. In spite of an extraordinary worldwide effort devoted to Machine Learning technology, the quality management of AI systems is fragmented and incomplete. In order to meet the needs of society, we need an AI engineering framework that meets the rigor needed. This paper provides an early glimpse of such a framework.  

\section*{Acknowledgements}
Author thanks Rachel Bellamy, Evelyn Duesterwald, Eitan Farchi, Michael Hind, David Porter, Orna Raz and Jim Spohrer for useful comments on the paper.

\bibliographystyle{splncs04}
\bibliography{mybibliography}

\end{document}